\documentclass{emulateapj}
\usepackage{amsmath}
\usepackage{array}
\usepackage{float}
\usepackage{graphicx}
\usepackage{subfigure}
\usepackage{natbib}

\begin{document}
\title{The color-magnitude distribution of small Jupiter Trojans \footnotemark[*]}
\author{Ian Wong and Michael E. Brown}
\affil{Division of Geological and Planetary Sciences, California Institute of Technology,
Pasadena, CA 91125, USA; iwong@caltech.edu}

\footnotetext[*]{Based on data collected at Subaru Telescope, which is operated by the National Astronomical Observatory of Japan.}

\begin{abstract}
We present an analysis of survey observations targeting the leading L4 Jupiter Trojan cloud near opposition using the wide-field Suprime-Cam CCD camera on the 8.2~m Subaru Telescope. The survey covered about 38~deg$^2$ of sky and imaged 147 fields spread across a wide region of the L4 cloud. Each field was imaged in both the $g'$ and the $i'$ band, allowing for the measurement of $g-i$ color. We detected 557 Trojans in the observed fields, ranging in absolute magnitude from $H=10.0$ to $H = 20.3$. We fit the total magnitude distribution to a broken power law and show that the power-law slope rolls over from $0.45\pm 0.05$ to $0.36^{+0.05}_{-0.09}$ at a break magnitude of $H_{b}=14.93^{+0.73}_{-0.88}$. Combining the best-fit magnitude distribution of faint objects from our survey with an analysis of the magnitude distribution of bright objects listed in the Minor Planet Center catalog, we obtain the absolute magnitude distribution of Trojans over the entire range from $H=7.2$ to $H=16.4$. We show that the $g-i$ color of Trojans decreases with increasing magnitude.  In the context of the less-red and red color populations, as classified in \citet{wong} using photometric and spectroscopic data, we demonstrate that the observed trend in color for the faint Trojans is consistent with the expected trend derived from extrapolation of the best-fit color population magnitude distributions for bright catalogued Trojans. This indicates a steady increase in the relative number of less-red objects with decreasing size. Finally, we interpret our results using collisional modeling and propose several hypotheses for the color evolution of the Jupiter Trojan population.
\end{abstract}

\section{Introduction}
Residing at a mean heliocentric distance of 5.2~AU, the Jupiter Trojans are asteroids that share Jupiter's orbit around the Sun and are grouped into two extended swarms centered around the stable L$_{4}$ and L$_{5}$ Lagrangian points. Estimates of the size of this population indicate that the Trojans are comparable in number to main belt asteroids of similar size \citep{szabo,nakamura}. Explaining the origin and evolution of this significant population of minor bodies is crucial for understanding the formation and dynamical history of the Solar System. While early theories posited that the Trojans could have formed out of the body of planetesimals and dust in the immediate vicinity of a growing Jupiter \citep{oldmodel}, later studies revealed that such \textit{in situ} formation is not consistent with the observed total mass and broad inclination distribution. An alternative theory suggests that the Trojans formed at large heliocentric distances out of the same body of material that produced the Kuiper Belt \citep{morbidelli}. Subsequent migration of the gas giants triggered a period of chaotic dynamical alterations in the outer Solar System, during which the primordial trans-Neptunian planetesimals were disrupted \citep{tsiganis,nesvorny}. It is hypothesized that a fraction of these objects were scattered inwards and captured by Jupiter as Trojan asteroids.

A detailed study of the size distribution of Trojans promises to shed light on the relationships between the Trojans and other minor body populations in the outer Solar System, and more broadly, constrain models of late Solar System evolution. The size distribution, or as a proxy, the magnitude distribution, offers significant insight into the nature of the Trojan population, as it contains information about the conditions in which the objects were formed as well as the processes that have shaped the population since its formation. Previous studies of the Trojan magnitude distribution have largely focused on objects larger than $\sim$10-20 km in diameter \citep[e.g.,][]{jewitt,szabo}, although the advent of larger telescopes and improved instruments has presented the opportunity to carry out surveys of smaller Trojans. \citet{yoshida} and  \citet{yoshida2} presented the first magnitude distribution for small L4 and L5 Trojans as part of a small survey, with several dozen objects detected down to sizes of $\sim$2 km. The detection of many more faint objects promises to expand our understanding of the small Trojan population.

Little is known about the composition and surface properties of Trojans. Both large-scale and targeted observational studies over the past few decades have revealed a population that is notably more homogeneous than the main belt asteroids, with low albedos and spectral slopes ranging form neutral to moderately red \citep[e.g.,][]{szabo,roig,fernandez}. Meanwhile, visible and near-infrared spectroscopy has been unable to detect any incontrovertible spectral features \citep[e.g.,][]{dotto,fornasier,yangjewitt,melita,emery}.  However, recent work has uncovered bimodalities in the distribution of various spectral properties, such as spectral slope in the visible \citep{szabo,roig,melita} and the near-infrared \citep{emery}. In \citet{wong}, the data from these previous studies were compiled and shown to be indicative of the existence of two color populations --- the so-called red and less-red populations. The magnitude distributions of these two populations are distinct, differing especially in the power-law distribution slopes of objects smaller than $\sim$50 km. Several hypotheses for the origin of this discrepancy have been posited, including different source regions for red and less-red Trojans, conversion of red objects to less-red fragments upon collision, and space weathering effects \citep{melita,wong}. By extending the analysis of Trojan colors to smaller objects, we hope to better understand the underlying processes behind the color dichotomy and the differing magnitude distributions of the color populations.

In this paper, we present the results of our survey of small Trojans in the leading L4 cloud. We detected over 550 Trojans and measured their brightness in two filters, from which their magnitudes and colors were computed. We calculate the best-fit curve describing the total magnitude distribution down to a limiting absolute magnitude of $H=16.4$. In addition, we present the first analysis of the color distribution of faint Trojans to date and compare the measured trends with previously-published results for brighter Trojans \citep{wong}. Lastly, we use collisional modeling to interpret the derived magnitude and color distributions.

\section{Observations}\label{sec:obs}

\begin{figure*}[t!]
\begin{center}
\includegraphics[width=17cm]{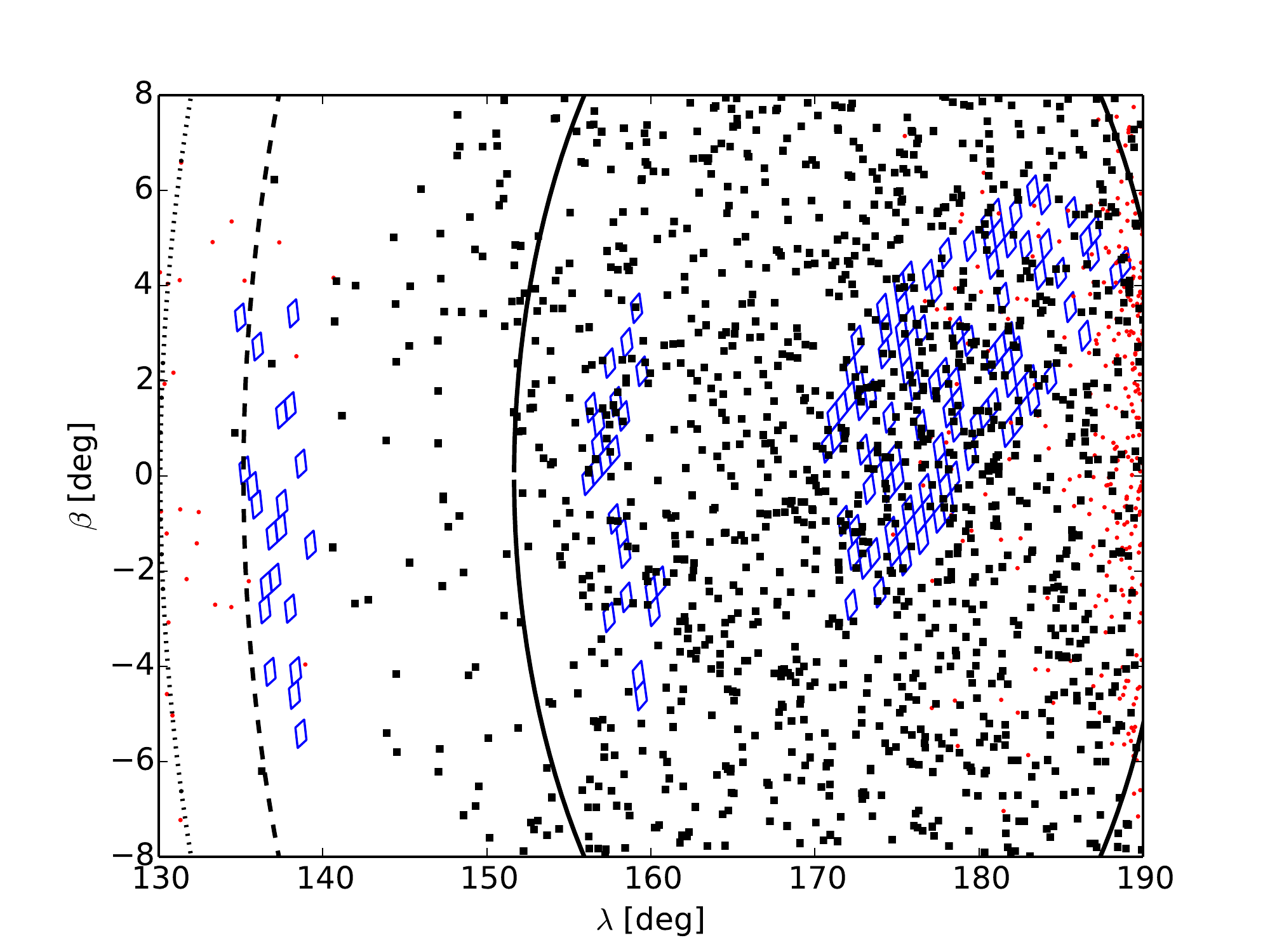}
\end{center}
\caption{Locations of the 147 observed Subaru Suprime-Cam fields, projected in geocentric ecliptic longitude-latitude space (blue diamonds). The size of the diamonds corresponds to the total field of view of each image. The positions of numbered Trojans and non-Trojans during the time of our observations with apparent sky motions in the range $14 \le |v| < 22{''/}\mathrm{hr}$ are indicated by black squares and red dots, respectively. The solid, dashed, and dotted curves denote respectively the approximate 50\%, 10\%, and 5\% relative density contours in the sky-projected L4 Trojan distribution \citep{szabo}.} \label{positions}
\end{figure*}

Observations of the L4 Trojan cloud were carried out on UT 2014 February 27 and 28 at the 8.2~m Subaru Telescope situated atop Mauna Kea, Hawaii. Using the Suprime-Cam instrument --- a mosaic CCD camera consisting of ten $2048\times4096$~pixel CCD chips that covers a $34'\times27'$ field of view with a pixel scale of $0.20''$ \citep{miyazaki} --- we observed 147 fields, corresponding to a total survey area of 37.5~deg$^{2}$. To detect moving objects as well as obtain color photometry, we imaged each field four times --- twice in the $g'$ filter ($\lambda_{\mathrm{eff}}=480.9$~nm) and twice in the $i'$ filter ($\lambda_{\mathrm{eff}}=770.9$~nm). The average time interval between epochs is about 20 minutes for images in the same filter, and about 30 minutes for images in different filters. We chose an exposure time of 60 seconds for all images to optimize survey depth and coverage. The resulting average observational arc for each moving object is roughly 70 minutes.

The on-sky positions of the 147 observed Suprime-Cam fields are shown in Figure~\ref{positions}. The surveyed region was divided into blocks of $10-12$ observing fields, which we imaged in the filter order $g'-g'-i'-i'$, or in reverse.  Blocks observed toward the beginning of each night targeted the trailing edge of the L4 Trojan cloud, while blocks observed later in the night are concentrated closer to the peak of the spatial distribution. To place the observed field locations in the context of the L4 point and the spatial extent of the leading Trojan cloud, we use the empirical L4 Trojan number density model from \citet{szabo}: $n(\lambda_{J}',\beta_{J}')\sim \exp{[-(\lambda_{J}'-60^{\circ})^{2}/2\sigma_{\lambda}^{2}}]\times\exp{[-\beta_{J}'^{2}/2\sigma_{\beta}^{2}}]$ with $\sigma_{\lambda}=14^{\circ}$ and $\sigma_{\beta}=9^{\circ}$, where $\lambda_{J}'$ and $\beta_{J}'$ are respectively the heliocentric ecliptic longitude and latitude relative to Jupiter. In Figure~\ref{positions}, the 50\%, 10\%, and 5\% relative number density contours are shown in geocentric longitude-latitude space $(\lambda,\beta)$; the position of the L4 point during the time of our observations is at around $(\lambda,\beta)=(170^{\circ},0^{\circ})$. We see that the majority of the observed fields lie in regions with predicted Trojan number densities at least 50\% of the peak value.

\begin{figure}[t!]
\begin{center}
\includegraphics[width=9cm]{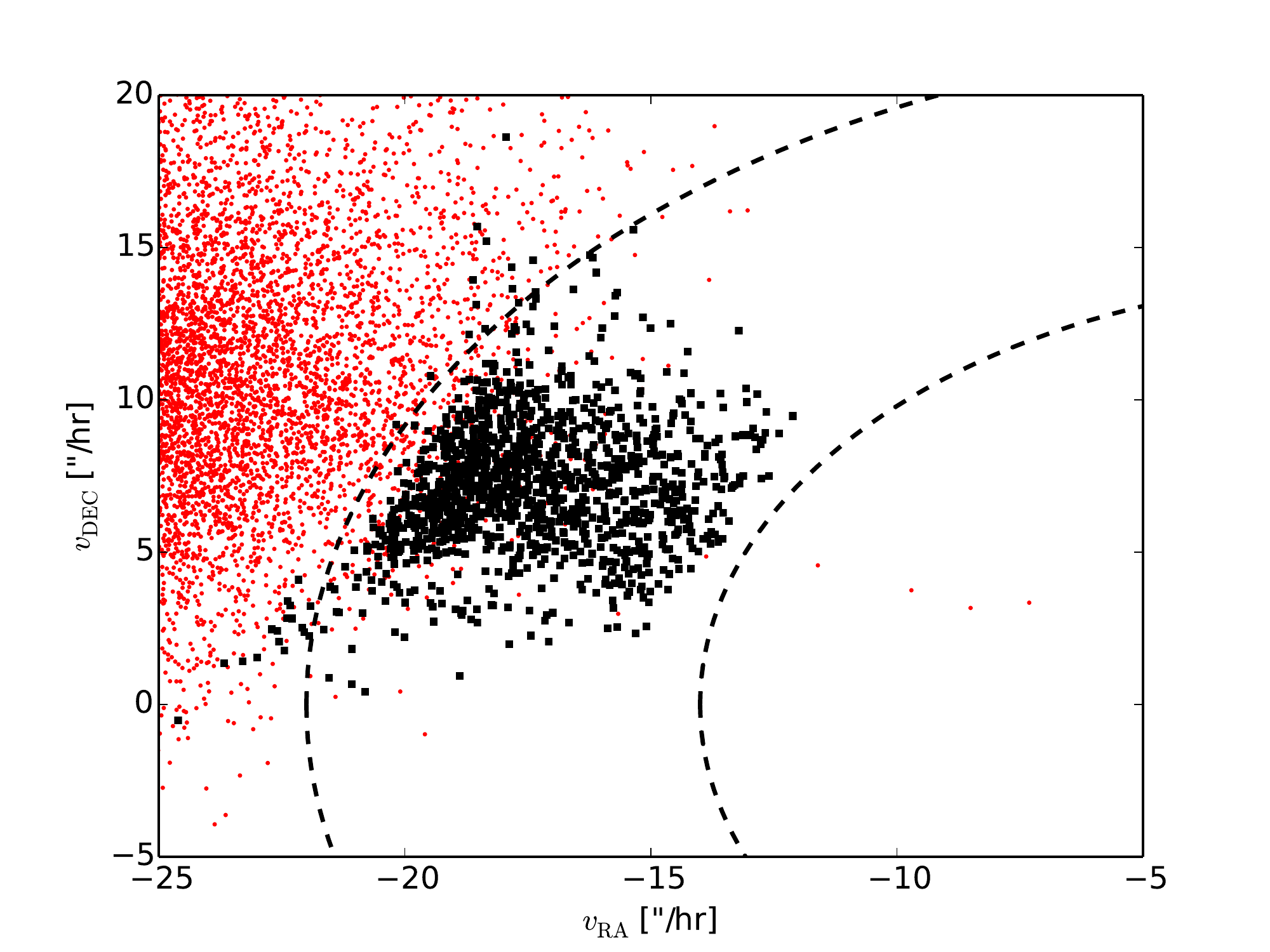}
\end{center}
\caption{Distribution of apparent RA and Dec velocities for numbered Trojans (black squares) and non-Trojans (red dots) with positions in the range $130^{\circ}\le \lambda < 190^{\circ}$ and $-8^{\circ}\le\beta < 8^{\circ}$. The dotted curves denote the $|v|=14''/\mathrm{hr}$ and $|v|=22''/\mathrm{hr}$ contours, which separate the Trojans from the majority of non-Trojan minor bodies.}\label{velocities}
\end{figure}

\subsection{Moving object detection}

After bias-subtracting the images, we flat-fielded them using the twilight flats taken at the start of the first night of observation. For every chip image (10 per exposure, 40 per observed field, 5880 for the entire survey), we calculated the astrometric solution by first creating a pixel position catalog of all the bright sources in the image. This was done using Version 2.11 of SExtractor \citep{sextractor}, with the threshold for source detection set at a high value (typically 30 or higher, depending on the seeing, in units of the estimated background standard deviation). The source catalog was then passed to Version 1.7.0 of SCAMP \citep{scamp}, which matches objects in the source catalog with those in a reference catalog of stars and computes the astrometric projection parameters for each chip image; we used the 9th Data Release of the Sloan Digital Sky Survey (SDSS DR9) as our reference catalog. In order to assess the quality of the astrometric solution from SCAMP, we compared the corrected on-sky position of stars in each image with the position of matched reference stars and found typical residual RMS values much less than $0.1''$. Likewise, we calculated the position scatter between matched stars from pairs of images taken in the same observing field and found typical RMS values less than $0.05''$.

Next, we passed the distortion-corrected images through SExtractor again, this time setting the detection threshold at 1.2 times the background standard deviation; we also required detected sources to consist of at least two adjacent pixels with pixel values above the detection threshold. These conditions were chosen to minimize the detection of faint non-astrophysical sources in the background noise as well as the loss of possible moving objects of interest. The resulting list contains the right ascension (RA) and declination (Dec) of all detected objects within each image. To search for moving object candidates, we fit orbits through sets of four source positions, one from each of the four images in an observing field, using the methods described in \citet{bernstein}. A source was flagged as a moving object candidate if the resulting $\chi^{2}$ value from the fit was less than 10. We reduced the number of non-Trojan moving object candidates flagged in this procedure by only considering sets of source positions consistent with apparent sky motion $|v|$ less than $25''/\mathrm{hr}$. For comparison, typical sky motions of known Trojans during the time of our observations lie in the range $14 \le |v| < 22{''/}\mathrm{hr}$.

Each moving object candidate was verified by aligning and blinking $100\times100$~pixel stamps clipped from each of the four images in the vicinity of the object. We rejected all non-asteroidal moving object candidates (e.g., sets of four sources that were flagged by the previous orbit-fitting procedure, but that included cosmic ray hits and/or anomalous chip artifacts). We also rejected asteroids that passed in front of background stars, coincided with a cosmic ray hit in one or more image, or otherwise traversed regions on the chip that would result in unreliable magnitude measurements. These objects numbered around 10, of which only 2 had on-sky positions and apparent sky motions consistent with Trojans (see Section~\ref{subsec:selection}). Therefore, the removal of these objects from our Trojan data set is not expected to have any significant effect on the results of our analysis. Through the procedures described above, we arrived at an initial set of 1149 moving objects.

\subsection{Selecting Trojans}\label{subsec:selection}

\begin{figure*}[t!]
\begin{center}
\includegraphics[width=17cm]{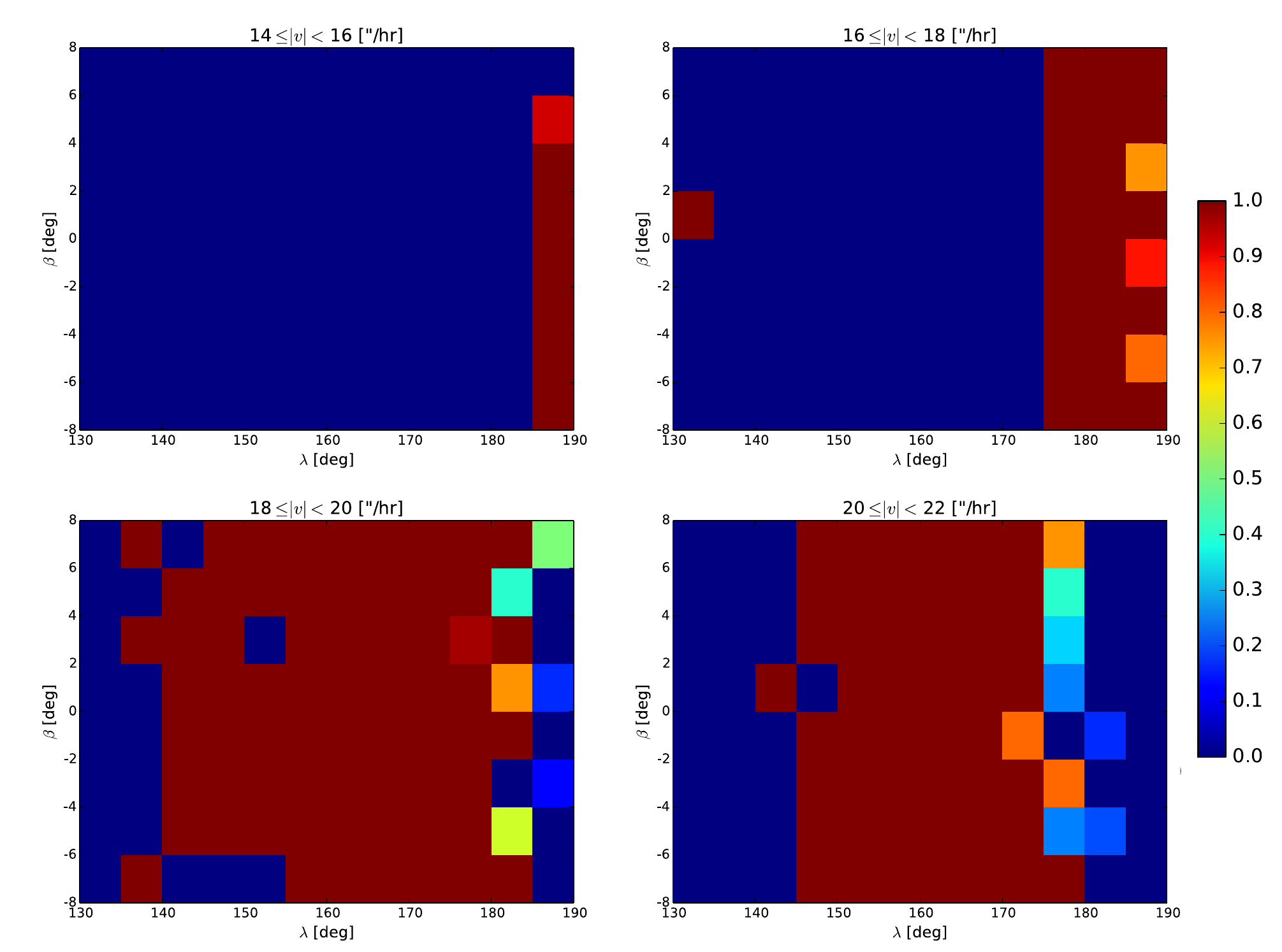}
\end{center}
\caption{Map of Trojan fraction $\gamma$ among numbered minor bodies at various locations in the space of ecliptic longitude, ecliptic latitude, and apparent sky motion during the time of our observations. Regions in dark red have 0\% predicted contamination by non-Trojans, and our operational Trojan data set includes only objects detected in observed fields located within these regions.} \label{filtering}
\end{figure*}

Observations were taken when the L4 point was near opposition, where the apparent sky motion of an object, $|v|$, is roughly inversely related to the heliocentric distance. Since the short observational arc of $\sim$70 minutes prevented an accurate determination of the heliocentric distance from orbit-fitting, we resorted to using primarily sky motion to distinguish Trojans from Main Belt Asteroids and Hildas. Figure~\ref{velocities} shows the distribution of apparent velocities in RA and Dec for all numbered minor bodies as calculated from ephemerides generated by the JPL HORIZONS system for UT 2014 February 27 12:00 (roughly the middle of our first night of observation). We have shown only objects with on-sky positions in the range of our observing fields ($130^{\circ}\le \lambda < 190^{\circ}$ and $-8^{\circ}\le\beta< 8^{\circ}$). A large majority of the objects with $|v|<20''/\mathrm{hr}$ are Trojans, while the relative portion of non-Trojan contaminants increases rapidly in the range $20''/\mathrm{hr}]\le |v|<22''/\mathrm{hr}$. We made an initial cut in angular velocity to consider only objects in the range $14\le |v|<22''/\mathrm{hr}$.

The contamination rate varies across the surveyed area as a function of apparent sky motion, as well as position on the sky. We carried out a more detailed contamination analysis by applying a rough grid in the space of $|v|$, $\lambda$, and $\beta$ spanning the range $14\le |v|<22''/\mathrm{hr}$, $130^{\circ}\le \lambda < 190^{\circ}$, and $-8^{\circ}\le\beta< 8^{\circ}$. We then binned all numbered objects contained in the JPL HORIZONS system into this three-parameter space and computed the Trojan fraction $\gamma\equiv$~[numbered Trojans]/[all numbered objects] in each bin. For bins containing no numbered objects, we assigned a value $\gamma=0$. Figure~\ref{filtering} shows the results of this analysis. We note that while the non-Trojan contamination rate over the whole surveyed area among objects with apparent sky motions between 20 and $22''/\mathrm{hr}$ is high, there are regions in the sky where the Trojan fraction is 1. Similarly, for objects with $|v|<20''/\mathrm{hr}$, there are regions near the edges of the ecliptic longitude space covered by our survey where the non-Trojan contamination rate is very high. 

In this paper, our operational Trojan data set includes only those objects detected in regions of the $(\lambda,\beta,|v|)$ space with Trojan fraction $\gamma=1$ (i.e., zero expected contamination). The resulting data set contains 557 Trojans. Choosing a slightly more lenient Trojan selection criterion (e.g., including objects detected in bins that do not contain numbered objects, where the Trojan fraction is technically unknown) was not found to significantly affect the main results of our magnitude and color distribution analysis.

\subsection{Photometric calibration}\label{subsec:photometry}

The apparent magnitude $m$ of an object detected in our images is given by $m = m_{0}-2.5\log_{10}(f) = m_{0}+m_{s}$, where $f$ is the measured flux, $m_{0}$ is the zero-point magnitude of the corresponding chip image, and the survey magnitude has been defined as $m_{s}\equiv -2.5\log_{10}(f)$. The flux of each object was calculated by SExtractor through a fixed circular aperture with a diameter of 5~pixels. We computed the zero-point magnitude for each chip image by matching all bright, non-saturated sources detected by SExtractor to reference stars in the SDSS DR9 catalog and fitting a line with slope one through the points $(m_{s},m)$. The maximum allowed position difference for a match between image and reference stars was set at $0.5''$, which resulted in an average of $\sim$150 matched stars per chip image. We used Sloan $g$-band and $i$-band reference star magnitudes to calibrate images taken in the $g'$ and $i'$ filters of Suprime-Cam, respectively. Consequently, the calculated apparent magnitudes of our detected Trojans are effectively Sloan $g$ and $i$ magnitudes, which greatly facilitates the comparison of our computed colors with those derived for Trojans in the SDSS Moving Object Catalog (see Section~\ref{subsec:colordist}).

Both the error in the zero-point magnitude $\sigma_{0}$ and the error in the measured flux $\sigma_{f}$ (reported by SExtractor) contribute to the error in the apparent magnitude $\sigma$, which we calculate using standard error propagation methods: $\sigma = \sqrt{\sigma_{0}^{2}+(2.5(\sigma_{f}/f)/\ln(10))^{2}}$. We set the $g$ magnitude of each Trojan detected in our survey to be the error-weighted mean of the apparent magnitudes calculated from the two $g'$ Suprime-Cam images, i.e., $g=\left(\sum_{k=1}^{2}m_{g,k}/\sigma_{g,k}^{2}\right)/\left(\sum_{k=1}^{2}1/\sigma_{g,k}^{2}\right)$, with the corresponding uncertainty in the $g$ magnitude defined by $\sigma_{g}^{2}=1/\left(\sum_{k=1}^{2}1/\sigma_{g,k}^{2}\right)$; the $i$ magnitude and uncertainty of each Trojan were defined analogously.

As mentioned previously in Section~\ref{subsec:selection}, orbit-fitting over the short observational arcs prevented us from precisely determining the orbital parameters of the detected objects. Therefore, we did not directly convert the apparent magnitudes to absolute magnitudes using the best-fit heliocentric distances. Instead, we considered the difference between apparent V-band magnitude and absolute magnitude $H$ of known Trojans, as computed by the JPL HORIZONS system for the time of our observations. By fitting a linear trend through the computed apparent sky motions $|v|$ as a function of the magnitude difference values, we obtained an empirical conversion between the sky motion and $V-H$. Since $V-H$ depends also on the viewing geometry, we divided the ecliptic longitude range $130^{\circ}\le \lambda < 190^{\circ}$ into $5^{\circ}$ bins and derived linear fits separately for known Trojans within each bin. We calculated the apparent V-band magnitudes of the detected Trojans from our survey using the empirical mean colors $g-r=0.55$ and $V-r = 0.25$ reported in \citet{szabo}. Then, we translated these $V$ values to $H$ using the measured sky motions and our empirical $V-H$ conversions. 

From the absolute magnitude, the diameter of a Trojan can be estimated using the relation $D = 1329\times 10^{-H/5}/\sqrt{p_{v}}$, where $D$ is in units of kilometers, and we assumed a uniform geometric albedo of $p_{v}=0.04$ \citep{fernandez}. The brightest Trojan we detected in our survey has $H=10.0$, corresponding to a diameter of 66.5~km, while the faintest Trojan has $H=20.3$, corresponding to a diameter of 0.6~km.

\subsection{Data completeness}\label{subsec:completeness}

\begin{figure}[t!]
\begin{center}
\includegraphics[width=9cm]{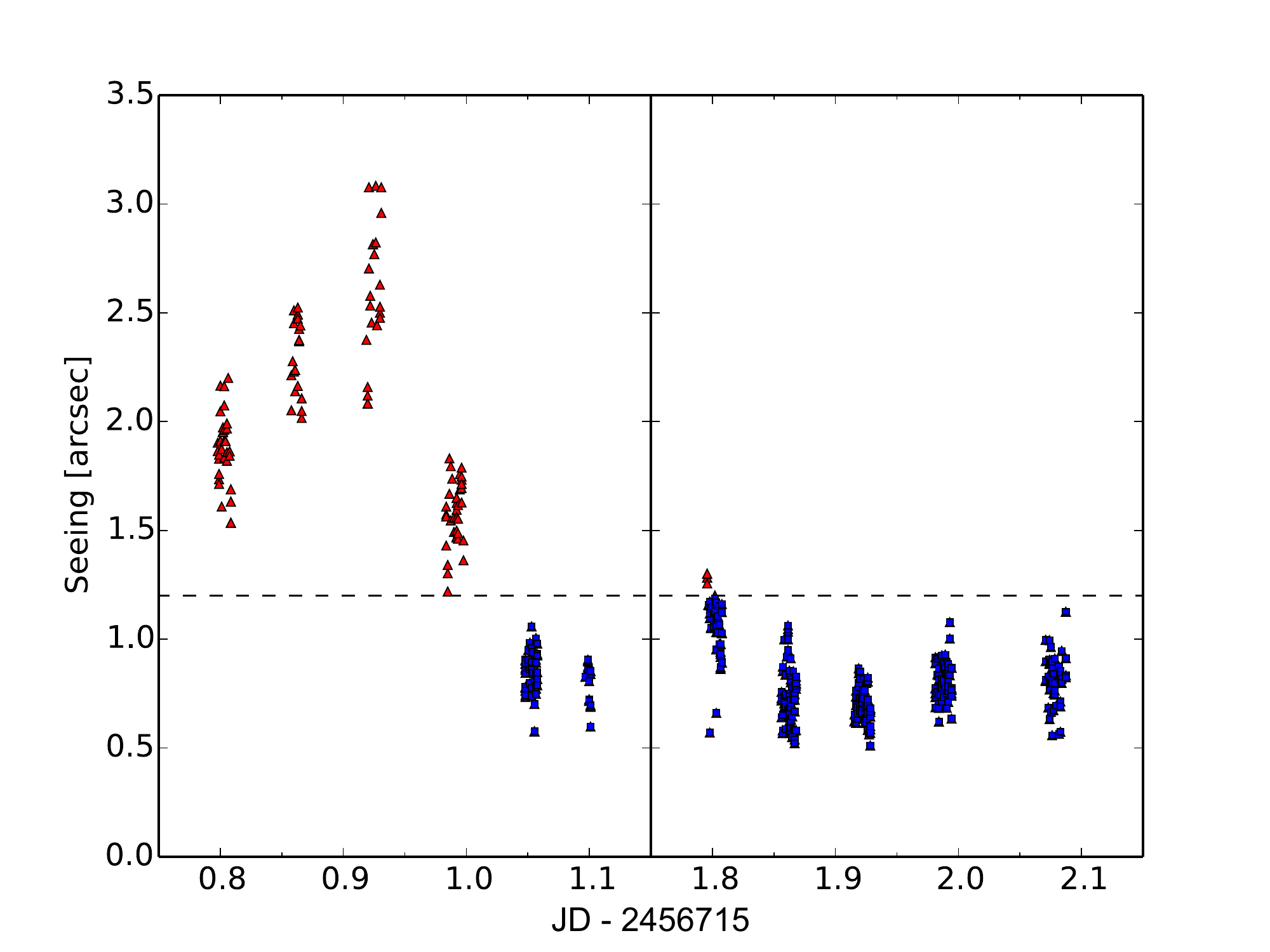}
\end{center}
\caption{Seeing of the worst image (highest seeing) in every chip for each set of four exposures in the 147 observed fields, plotted as a function of the first exposure time. The dotted line indicates the cutoff at $1.2''$ seeing - objects in fields with higher seeing (red triangles) are not included in the filtered Trojan set.}\label{seeing}
\end{figure}

An analysis of the magnitude distribution of a population can be severely affected by detection incompleteness. Varying weather conditions during ground-based surveys lead to significantly different detection completeness limits for observations taken at different times, resulting in a non-uniform data set and biases in the overall magnitude distribution. Since we required a moving object candidate to be detected in all four images taken in an observed field, the epoch with the highest seeing in each set of four exposures determined the threshold magnitude to which the survey was sensitive in that particular field. Here we set the seeing value of each chip image to be the median of the full-width half-max of the point-spread function for all non-saturated stars, as computed by SExtractor. If the highest seeing value across a four-image set was high, then the object detection pipeline would have missed many of the fainter objects positioned within the field. This is because the signal-to-noise of objects of a given magnitude located on the worst image of the set would be significantly lower than that of the same objects located on an image taken at good seeing. As a result, the magnitude distribution of objects detected in an observing field with high-seeing images would be strongly biased toward brighter objects, and when combined with the full body of data, would affect the overall magnitude distribution.

Figure~\ref{seeing} shows the highest measured seeing for each group of four chip images taken in an observing field, plotted with respect to the time of first epoch. A large portion of our first night of observation was plagued by very high seeing - as high as $3.0''$ at times. We chose a cutoff seeing value of $1.2''$ and defined a filtered Trojan set containing only Trojans detected in images with seeing below this value.

We used signal-to-noise to establish a limiting magnitude for our magnitude distribution analysis. Instead of attempting to generate an empirical model of the detection efficiency for the faintest objects, we elected to stipulate a conservative $S/N$ threshold of 8. We defined the upper magnitude limit of our filtered Trojan data set to be the magnitude of the brightest object (detected at seeing less than $1.2''$) with $S/N<8$. The limiting magnitude was determined to be $H=16.4$. The final filtered Trojan set contains 150 objects, and when fitting for the total magnitude distribution, we assume that this set is complete (i.e., is not characterized by any size-dependent bias).

\subsection{Trojan colors}\label{subsec:colors}

The color $c$ of each Trojan is defined as the difference between the $g$ and $i$ magnitudes: $c\equiv g-i$. When calculating the uncertainty of each color measurement, we must consider the effect of asteroid rotation in addition to the contribution from photometric error. The oscillations in apparent amplitude seen in a typical asteroidal rotational light curve arise from the non-spherical shape of the object and peak twice during one full rotation of the asteroid \citep[see, for example, the review by][]{pravec}. The average observational arc of $\sim$70 minutes for each object may correspond to a significant fraction of a characteristic light curve oscillation period, which can consequently lead to a large variation in the apparent brightness of an object across the four epochs. 

\begin{figure}[t!]
\begin{center}
\includegraphics[width=9cm]{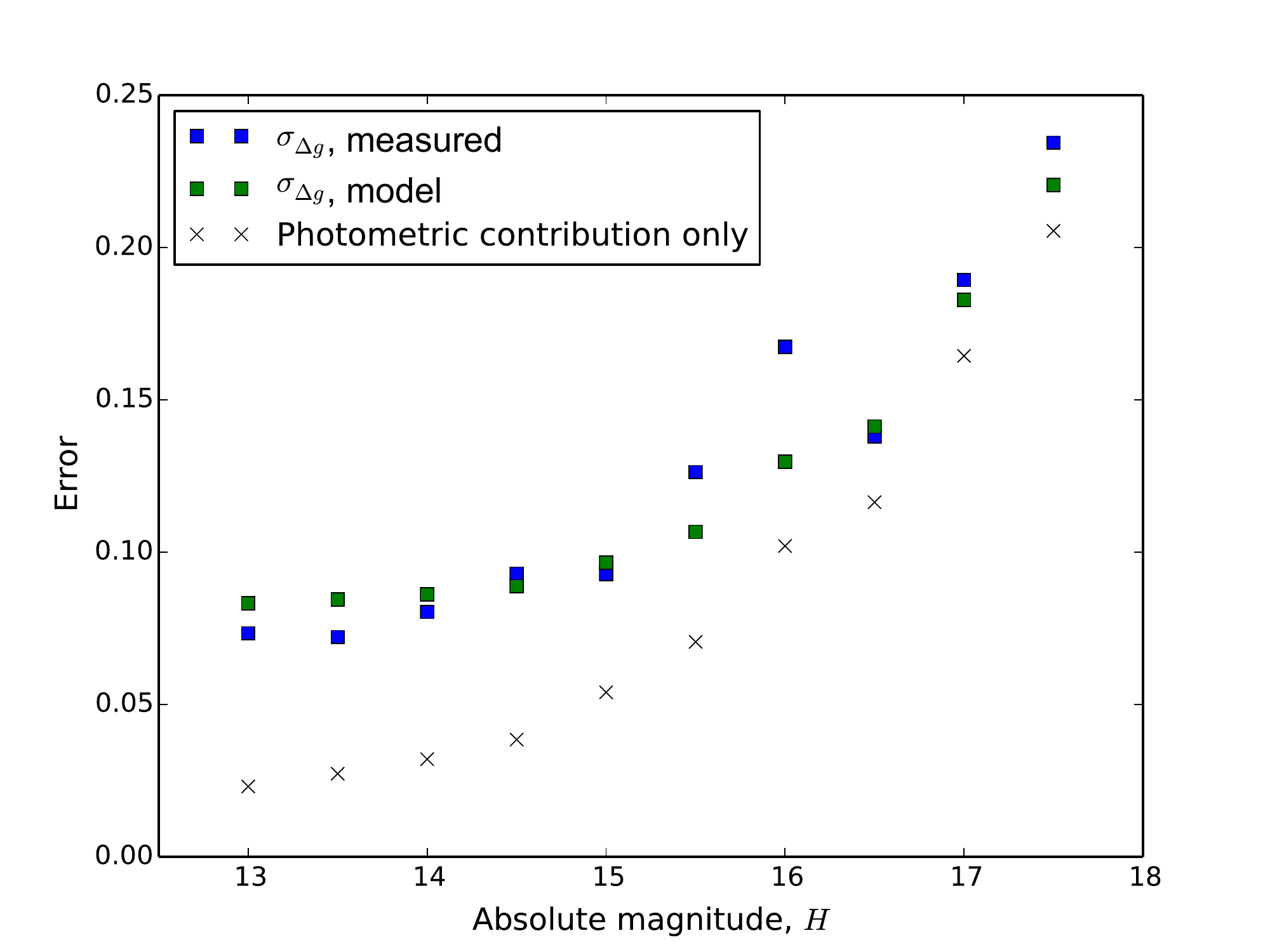}
\end{center}
\caption{Comparison of the measured standard deviation error in the difference of $g$ magnitudes, $\sigma_{\Delta g}$, binned in 0.5~mag intervals (blue squares) with the corresponding values from our empirical model combining the measured photometric magnitude errors with a constant contribution from asteroid rotation (green squares). The binned medians of the photometric errors only are denoted by black crosses. The agreement between the measured and modeled $\sigma_{\Delta g}$ values shows that the assumption of a magnitude-invariant contribution to the dispersion in $\Delta g$ from asteroid rotation is good.}\label{rotate}
\end{figure}

In the absence of published light curves for faint Trojans, we must develop a model of the rotational contribution to the color uncertainty in order to accurately determine the total uncertainty of our color measurements. We took advantage of the fact that we obtained two detections of an object in each filter to estimate the effect of asteroid rotation empirically. We considered the difference between the two consecutive $g'$-band magnitude measurements $\Delta g \equiv g_{1}-g_{2}$. The standard deviation in $\Delta g$ values, $\sigma_{\Delta g}$, contains a contribution from the photometric error given by the quadrature sum of the individual magnitude uncertainties as defined in Section~\ref{subsec:photometry}: $\sigma_{\Delta g,\mathrm{phot}}= \sqrt{\sigma_{g,1}^{2}+\sigma_{g,2}^{2}}$. Figure~\ref{rotate} compares the standard deviation of $\Delta g$ values measured in 0.5~mag bins (blue squares) to the corresponding error contribution from the photometric uncertainty only, $\sigma_{\Delta g,\mathrm{phot}}$ (black crosses). 

From the plot, it is evident that an additional contribution attributable to rotation is needed to account for the uncertainty in $\Delta g$ seen in our data. We observe that the discrepancy between $\sigma_{\Delta g}$ and the photometric error contribution is large at low magnitudes, where the photometric error is small, and decreases with increasing magnitude and increasing photometric error. This overall trend suggests that the total error in $\Delta g$ can be empirically modeled as a quadrature sum of the photometric contribution and the rotational contribution, i.e., $\sigma_{\Delta g}=\sqrt{\sigma_{\Delta g, \mathrm{phot}}^{2}+\sigma_{\Delta g, \mathrm{rot}}^{2}}$, where the rotational contribution is magnitude-invariant. We found that a rotational contribution of $\sigma_{\Delta g, \mathrm{rot}}\sim 0.08$ gives a good match with the measured standard deviation in $\Delta g$, and likewise for $\Delta i$. In Figure~\ref{rotate}, the green squares show the binned $\Delta g$ standard deviation values calculated from our empirical model combining both photometric and rotational contributions.

The rotational contribution to the color uncertainty was estimated as a quadrature sum of the rotational contributions to $\Delta g$ and $\Delta i$: $\sigma_{c,\mathrm{rot}}\sim0.11$~mag. The total color uncertainty is therefore given by:
\begin{equation}\label{colorerr}\sigma_{c} = \sqrt{\sigma_{g}^{2}+\sigma_{i}^{2}+\sigma_{c,\mathrm{rot}}^{2}},\end{equation}
where $\sigma_{g}$ and $\sigma_{i}$ are the errors in the $g$ and $i$ magnitudes derived from the flux and calibrated zero-point magnitude errors only (see Section~\ref{subsec:photometry}).

We note that the effects of bad seeing and detection incompleteness discussed in Section~\ref{subsec:completeness} are not expected to affect the bulk properties of the Trojan color distribution (e.g., mean color), except at the faintest magnitudes, where there is a slight bias toward redder objects. This is because we used the measured $g$ magnitudes to convert to absolute magnitudes $H$. For objects with a given $g$ magnitude, redder objects are slightly brighter in the $i'$-band (lower $i$ magnitude) than less-red objects. The predicted effect is small, since the characteristic difference in $g-i$ color between objects in the red and less-red populations is only $\sim$0.15~mag (see Section~\ref{subsec:colordist}). In our color analysis, we strove to avoid this small color bias by limiting the analysis to objects with $H$ magnitudes less than 18, thereby removing the faintest several dozen objects from consideration.

\section{Analysis}
In this section, we first present our analysis of the total Trojan magnitude distribution, using both data for faint Trojans from our Subaru survey and data for brighter L4 Trojans listed in the Minor Planet Center catalog. Next, the distribution of measured $g-i$ colors and its magnitude dependence studied in the context of the less-red and red color populations described in \citet{wong}.

\subsection{Total magnitude distribution}\label{subsec:totaldist}

\begin{figure}[t!]
\begin{center}
\includegraphics[width=9cm]{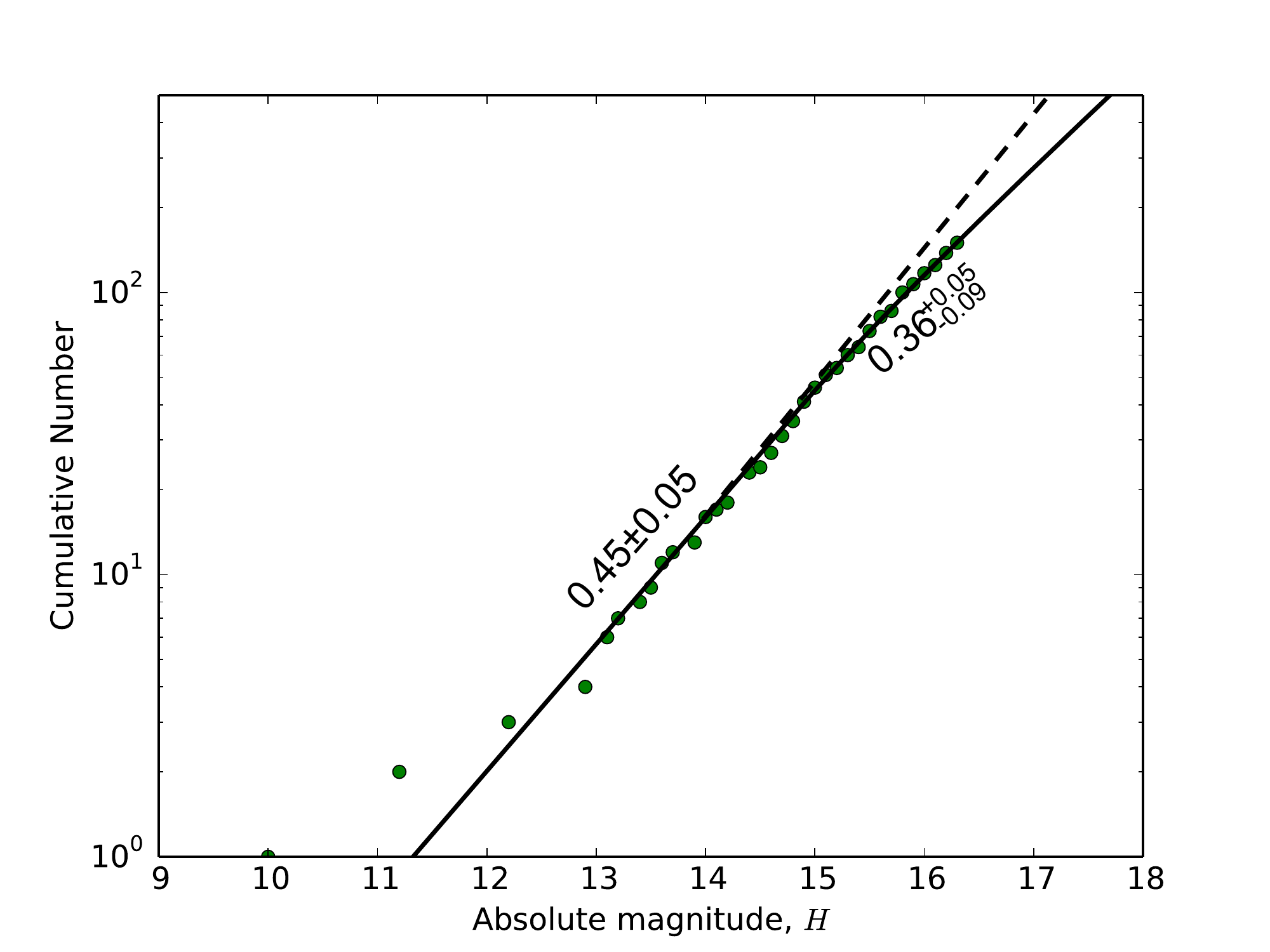}
\end{center}
\caption{Cumulative absolute magnitude distribution of the filtered Trojan set from our Subaru observations, binned by 0.1~mag (green dots). The best-fit broken power law curve describing the distribution is overplotted (solid black line), with the power law slopes indicated. The dashed line is an extension of the $\alpha_{1}=0.45$ slope and is included to make the slope rollover more discernible.}\label{cumulative}
\end{figure}

The cumulative magnitude distribution $N(H)$ of all objects in the filtered Trojan set (Section~\ref{subsec:completeness}) as a function of $H$ magnitude is shown in Figure~\ref{cumulative}. The main feature of the magnitude distribution is a slight rollover in slope to a shallower value at around $H\sim15$. Following previous analyses of the magnitude distribution of Trojans \citep[e.g.,][]{jewitt}, we fit the differential magnitude distribution, $\Sigma(H) = dN(M)/dH$, with a broken power law
\begin{align}\label{distribution}\Sigma(\alpha_{1},\alpha_{2},&H_{0},H_{b}|H) \notag\\
&= 
\begin{cases}
10^{\alpha_{1}(H-H_{0})},&\text{for }H< H_{b}\\
10^{\alpha_{2}H+(\alpha_{1}-\alpha_{2})H_{b} - \alpha_{1}H_{0}},&\text{for }H \ge H_{b}
\end{cases},\end{align}
where the power law slope for brighter objects $\alpha_{1}$ changes to a shallower faint-end slope $\alpha_{2}$ at a break magnitude $H_{b}$. The parameter $H_{0}$ defines the threshold magnitude for which $\Sigma(H_{0}) =1$.

We fit the model curve in Eq.~\eqref{distribution} to our filtered Trojan magnitude distribution using a maximum likelihood method similar to the one used in \cite{fraser} for their study of Kuiper belt objects. We defined a likelihood function $L$ that quantifies the probability that a random sampling of the model distribution will yield the data:
\begin{equation}\label{likelihood}L(\alpha_{1},\alpha_{2},H_{0},H_{b}| H_{i})\propto e^{-N}\prod_{i}P_{i}.\end{equation}
Here, $H_{i}$ is the $H$ magnitude of each detected Trojan and $P_{i}=\Sigma(\alpha_{1},\alpha_{2},H_{0},H_{b}|H_{i})$ is the probability of detecting an object $i$ with magnitude $H_{i}$ given the underlying distribution function $\Sigma$. $N$ is the total number of objects detected in the magnitude range under consideration and is given by
\begin{equation}\label{N}N = \int^{H_{\mathrm{max}}}_{-\infty} \eta(H)\Sigma(\alpha_{1},\alpha_{2},H_{0},H_{b}|H) \,dH,\end{equation}
where in the case of our filtered Trojan set we have $H_{\mathrm{max}}=16.4$. We have included the so-called ``efficiency'' function $\eta(H)$ that represents an empirical model of the incompleteness in a given data set and ensures that the best-fit distribution curve corrects for any incompleteness in the data. Our filtered Trojan data set was defined to remove incompleteness contributions from both bad seeing and low signal-to-noise, so when fitting the broken power law to the corresponding magnitude distribution, we set $\eta=1$.

We used an affine-invariant Markov Chain Monte Carlo (MCMC) Ensemble sampler \citep{mcmc} with 50,000 steps to estimate the best-fit parameters and corresponding 1$\sigma$ uncertainties. The magnitude distribution of the filtered Trojan set is best-fit by a broken power law with parameter values $\alpha_{1}=0.45\pm 0.05$, $\alpha_{2}=0.36^{+0.05}_{-0.09}$, $H_{0}=11.39^{+0.31}_{-0.37}$, and $H_{b}=14.93^{+0.73}_{-0.88}$. The cumulative magnitude distribution for this best-fit model is plotted in Figure~\ref{cumulative} as a solid black line. We note that while the difference between the two power-law slopes is small, it is statistically significant: Marginalizing over the slope difference $\Delta \alpha = \alpha_{1}-\alpha_{2}$, we obtained $\Delta \alpha = 0.10^{+0.09}_{-0.08}$, which demonstrates that the difference between the two power-law slopes is distinct from zero at the $1.25\sigma$ level.

\begin{figure}[t!]
\begin{center}
\includegraphics[width=9cm]{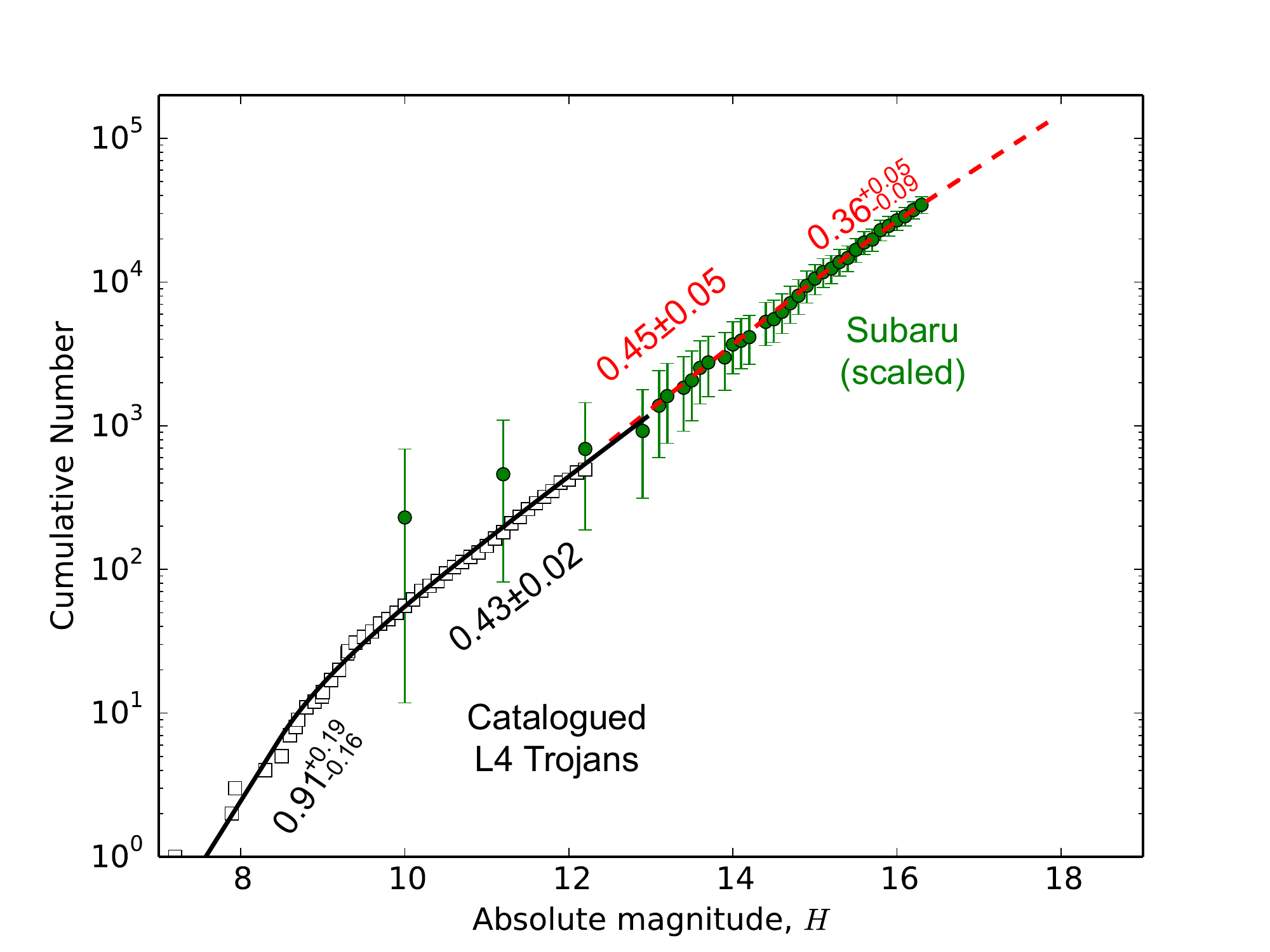}
\end{center}
\caption{Cumulative magnitude distribution of all L4 Trojans contained in the Minor Planet Center (MPC) catalog brighter than $H=12.3$ \citep[white squares; corrected for incompleteness in the range $H = 11.3-12.3$ following the methodology of][]{wong} and the cumulative magnitude distribution of the filtered Trojan set from our Subaru observations, approximately scaled to reflect the true overall number and binned by 0.1~mag (green dots). The error bars on the scaled Subaru Trojan magnitude distribution (in green) denote the 95\% confidence bounds derived from scaling the Poisson errors on the Subaru survey data. The uncertainties on the corrected MPC catalog Trojan magnitude distribution in the range $H = 11.3-12.3$ are much smaller than the points. The best-fit broken power law curves describing the MPC and Subaru data are overplotted (solid black and dashed red lines, respectively), with the power law slopes (in the text: $\alpha'_{0}$, $\alpha'_{1}$, $\alpha_{1}$, and $\alpha_{2}$) indicated for their corresponding magnitude regions.}\label{total}
\end{figure}

We also fit the magnitude distribution of known bright Trojans contained in the Minor Planet Center (MPC) catalog by repeating the analysis described in \citet{wong}, this time including only L4 Trojans. Fitting the distribution of bright L4 Trojans likewise to a broken power law of the form shown in Eq.~\eqref{distribution}, we obtained a best-fit distribution function with $\alpha'_{0}=0.91^{+0.19}_{-0.16}$, $\alpha'_{1}=0.43\pm 0.02$, $H'_{0}=7.22^{+0.24}_{-0.25}$, and $H'_{b}=8.46^{+0.49}_{-0.54}$. We have denoted the best-fit parameters for the MPC Trojan magnitude distribution with primes to distinguish them from the best-fit parameters for the Subaru Trojans; the subscripts on the slope parameters $\alpha$ are adjusted to reflect the magnitude ranges they correspond to in the overall distribution. The cumulative magnitude distribution of MPC L4 Trojans through $H=12.3$ \citep[corrected for catalog incompleteness following the methodology described in][]{wong} is shown in Figure~\ref{total} along with the best-fit curve. 

Earlier studies of L4 Trojans in this size range have reported power law slopes that are consistent with our values: \citet{yoshida2} fit MPC L4 Trojans with magnitudes in the range $9.2 < H < 12.3$ and derived a slope of $0.40\pm0.02$. We note that their fit did not take into account the incompleteness in the MPC catalog, which we estimated in \citet{wong} to begin at $H=11.3$. Therefore, the slope value in \citet{yoshida2} is somewhat underestimated. \citet{jewitt} carried out a survey of L4 Trojans and computed slopes of $0.9\pm 0.2$ and $0.40\pm0.06$ over size ranges corresponding to our reported $\alpha'_{0}$ and $\alpha'_{1}$ slopes, respectively. Our best-fit power law slopes calculated for bright MPC L4 Trojans are consistent with these previously-published values at better than the $1\sigma$ level.

We combined the magnitude distributions of faint Subaru Trojans and brighter catalogued Trojans to arrive at the overall magnitude distribution of L4 Trojans, shown in Figure~\ref{total}, where the cumulative absolute magnitude distribution of faint Trojans was approximately scaled to match the overall number of MPC Trojans at $H\sim 12$. The error bars indicate 95\% confidence bounds derived from Poisson errors on the Subaru survey data and reflect the uncertainty associated with scaling up the survey magnitude distribution to approximate the magnitude distribution of the total L4 population. The combined Subaru and MPC data sets cover the entire magnitude range from $H=7.2$ to a limiting magnitude of $H_{\mathrm{max}}=16.4$. The best-fit power law distribution slopes in the region containing the overlap between the MPC and Subaru data sets ($\alpha'_{1}=0.43\pm 0.02$ and $\alpha_{1}=0.45\pm 0.05$) are statistically equivalent, indicating that the magnitude distribution between $H\sim 10$ and $H\sim 15$ is well-described by a single power law slope. The overall magnitude distribution is characterized by three distinct regions: The brightest L4 Trojans have a power law magnitude distribution with slope $\alpha_{0}=0.91$. At intermediate sizes, the magnitude distribution rolls over to a slope of $\alpha_{1}\sim 0.44$. Finally, the faintest objects detected in our Subaru survey are characterized by an even shallower magnitude distribution slope of $\alpha_{2}=0.36$. These three regions are separated at the break magnitudes $H'_{b}=8.46^{+0.49}_{-0.54}$ and $H_{b}=14.93^{+0.73}_{-0.88}$, which correspond to Trojans of size $135^{+38}_{-27}$~km and $7^{+3}_{-2}$~km, respectively.

In a previous study, \citet{yoshida} detected 51 faint L4 Trojans near opposition using the Suprime-Cam instrument and found the magnitude distribution to be well-described by a broken power law with a break at around $H\sim 16$ separating a brighter-end slope of $0.48\pm 0.02$ from a shallower faint-end slope of $0.26\pm 0.02$. The methods used in the \citet{yoshida} fits differed from ours in several ways: The data was binned prior to fitting, and the two slopes were determined from independent fits of the bright and faint halves of their data. To better compare the distribution of Trojans studied by \citet{yoshida} with the best-fit distribution we derived from our Subaru survey, we reanalyzed their data using the techniques described in this paper. Fitting all of the \citet{yoshida} magnitudes through $H=17.9$ (90\% completeness limit) to a broken power-law, we obtained the two slopes $\alpha_{1,\mathrm{Y\&N}}=0.44^{+0.07}_{-0.06}$ and $\alpha_{2,\mathrm{Y\&N}}=0.26^{+0.06}_{-0.04}$ and a roll-over at $H_{b,\mathrm{Y\&N}}=15.11^{+0.89}_{-1.02}$. These values are consistent with the corresponding best-fit values for $\alpha_{1}$, $\alpha_{2}$, and $H_{b}$ from the analysis of our Subaru survey data at better than the $1\sigma$ level. 

\subsection{Color distribution}\label{subsec:colordist}

Previous spectroscopic and photometric studies of Trojans have noted bimodality in the distribution of various properties, including the visible \citep{szabo,roig,melita} and near-infrared \citep{emery} spectral slope, as well as the infrared albedo \citep{grav}. \citet{wong} demonstrated that the bimodal trends were indicative of  two separate populations within the Trojans, which are referred to as the less-red and red populations, in accordance with their relative colors. It was further shown that the magnitude distributions of these two color populations are distinct, with notably different power-law slopes in the magnitude range $H\sim 9.5-12.3$.

\begin{figure}[t!]
\begin{center}
\includegraphics[width=9cm]{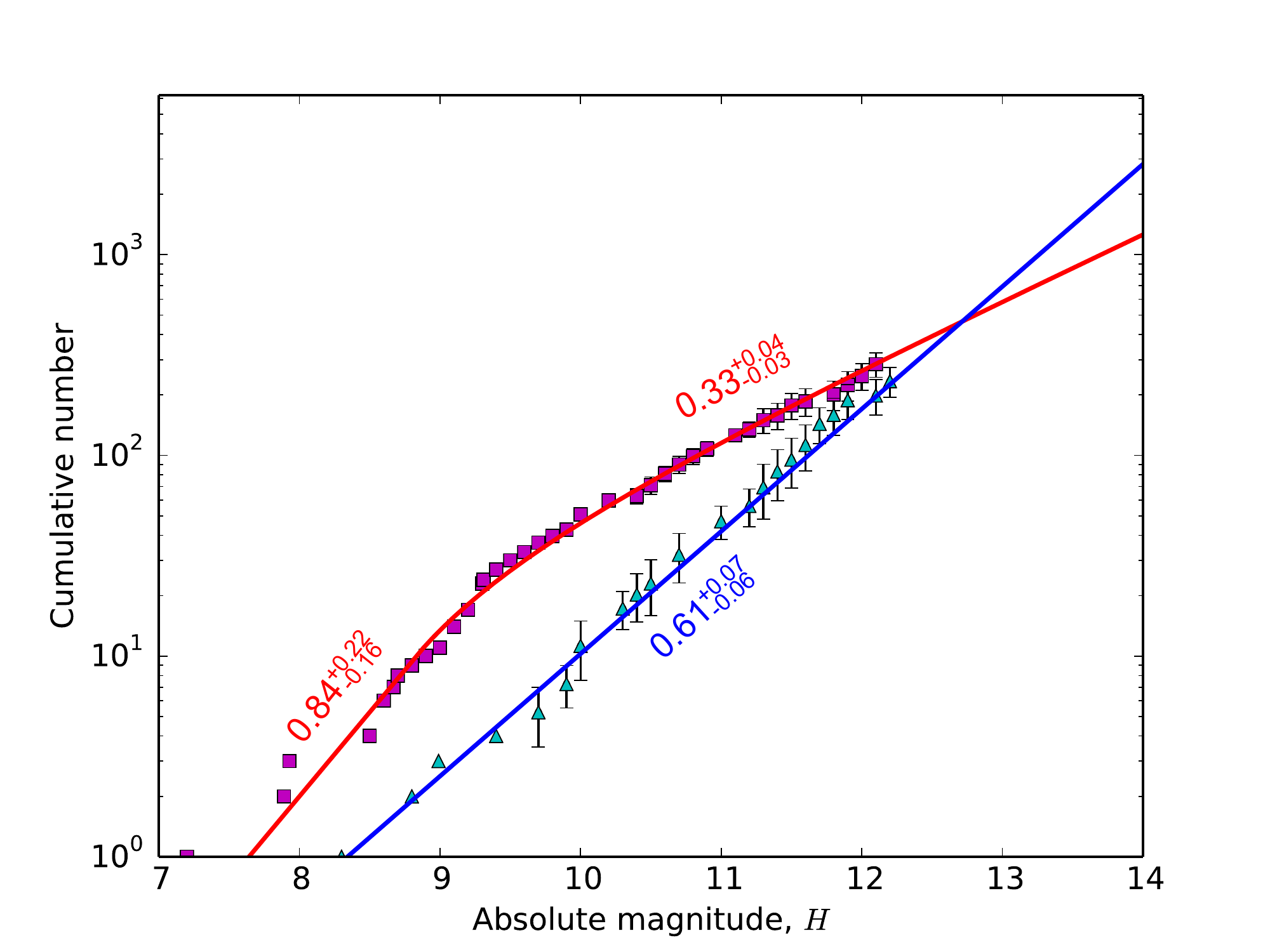}
\end{center}
\caption{Cumulative magnitude distribution of the less-red and red L4 Trojan populations, as constructed using the methods of \citet{wong} for objects in the Minor Planet Center catalog through $H=12.3$ (cyan triangles and red squares, respectively). These distributions have been scaled to correct for catalog and categorization incompleteness; the error bars denote the 95\% confidence bounds and are derived from the binomial distribution errors associated with correcting for uncategorized less-red and red Trojans, as well as the uncertainties from the catalog incompleteness correction in the range $H=11.3-12.3$. The best-fit curves describing the distributions are overplotted (solid blue and red lines, respectively), with the power law slopes indicated for their corresponding magnitude regions.}\label{colordist}
\end{figure}

We repeated the color population analysis presented in \citet{wong}, using only L4 Trojans. Objects were categorized as less-red or red primarily based on their visible spectral slopes, which were derived from the $g$, $r$, $i$, and $z$ magnitudes listed in 4th release of the SDSS Moving Object Catalog (SDSS-MOC4). We estimated the incompleteness in color categorization via the fraction of objects in each 0.1~mag bin that we were able to categorize as either less-red or red; when fitting the magnitude distributions of the color populations up to a limiting magnitude of $H=12.3$, the categorization incompleteness was factored into the efficiency function $\eta$ along with the estimated incompleteness of the MPC catalog. See \citet{wong} for a complete description of the methodology used.

The magnitude distributions of the color populations were fit using the same techniques that we applied in the total magnitude distribution fits in Section~\ref{subsec:totaldist}. The red population magnitude distribution is best-fit by a broken power law with $\alpha^{R}_{1}=0.84^{+0.22}_{-0.16}$, $\alpha^{R}_{2}=0.33^{+0.04}_{-0.03}$, $H^{R}_{0}=7.30^{+0.30}_{-0.26}$, and $H^{R}_{b}=8.82^{+0.35}_{-0.44}$. The less-red population has a magnitude distribution more consistent with a single power law: $\Sigma(\alpha_{1},H_{0}|H) = 10^{\alpha_{1}(H-H_{0})}$; here we computed the following best-fit parameters: $\alpha^{LR}_{1}=0.61^{+0.07}_{-0.06}$ and $H^{LR}_{0}=8.18^{+0.31}_{-0.29}$. Figure~\ref{colordist} shows the cumulative magnitude distributions for the less-red and red populations, scaled to correct for incompleteness, along with the best-fit curves. The error bars indicate the 95\% confidence bounds derived from the incompleteness correction. The best-fit slopes calculated above for the L4 color populations are consistent within the errors to the corresponding values in \citet{wong} derived from the color analysis of both L4 and L5 Trojans.

\begin{figure}[b!]
\begin{center}
\includegraphics[width=9cm]{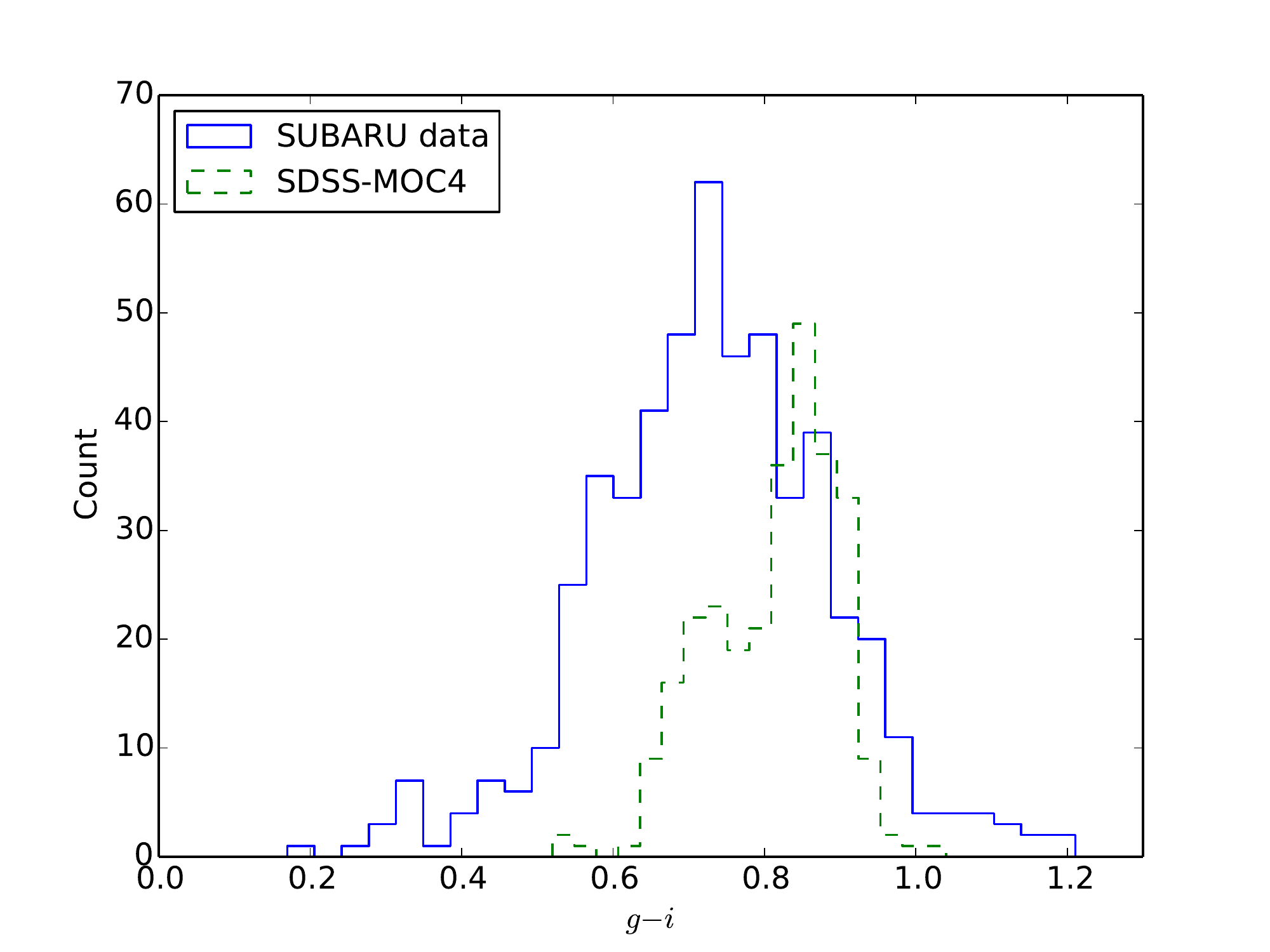}
\end{center}
\caption{Histogram of the $g-i$ color distribution for Trojans contained in the SDSS-MOC4 catalog brighter than $H=12.3$ (green) and fainter Trojans detected in our Subaru survey (blue). There is clear bimodality in the distribution of brighter objects in the SDSS-MOC4 catalog, while the distribution of faint Trojan colors does not display a clear bimodality. This is likely due to the large relative uncertainties associated with the measurement of faint Trojan colors due to asteroid rotation.}\label{colorhist}
\end{figure}

To determine whether the previously-studied bimodality in color among the brighter Trojans carries through to the fainter Trojans from our Subaru observations, we constructed a histogram of the $g-i$ color distribution for all Trojans contained in the SDSS-MOC4 catalog and compared it to the histogram of $g-i$ colors for Trojans detected in our Subaru survey. As discussed in Section~\ref{subsec:completeness}, the bulk distribution of colors is not expected to be affected by detection incompleteness due to bad seeing, and we include all Trojans brighter than $H=18$. The two histograms are plotted in Figure~\ref{colorhist}. The bimodality in the color histogram of brighter SDSS-MOC4 objects is evident. We fit a two-peaked Gaussian to the color distribution of brighter SDSS-MOC4 Trojans and found that the less-red and red populations have mean $g-i$ colors of $\mu_{1}=0.73$ and $\mu_{2}=0.86$, respectively.

On the other hand, while there is some asymmetry in the color distribution of faint Trojans detected by our Subaru survey, there is no robust bimodality. This can be mostly attributed to the large contribution of asteroid rotation to the variance in the color measurements, the magnitude of which is comparable to the difference between the mean red and less-red colors (see Section~\ref{subsec:colors}). As such, it is not possible to categorize individual Trojans from our Subaru observations into the less-red and red populations based on their colors and construct magnitude distributions of the color populations, as was done in \citet{wong} for the brighter SDSS-MOC4 Trojans.

\begin{figure}[b!]
\begin{center}
\includegraphics[width=9cm]{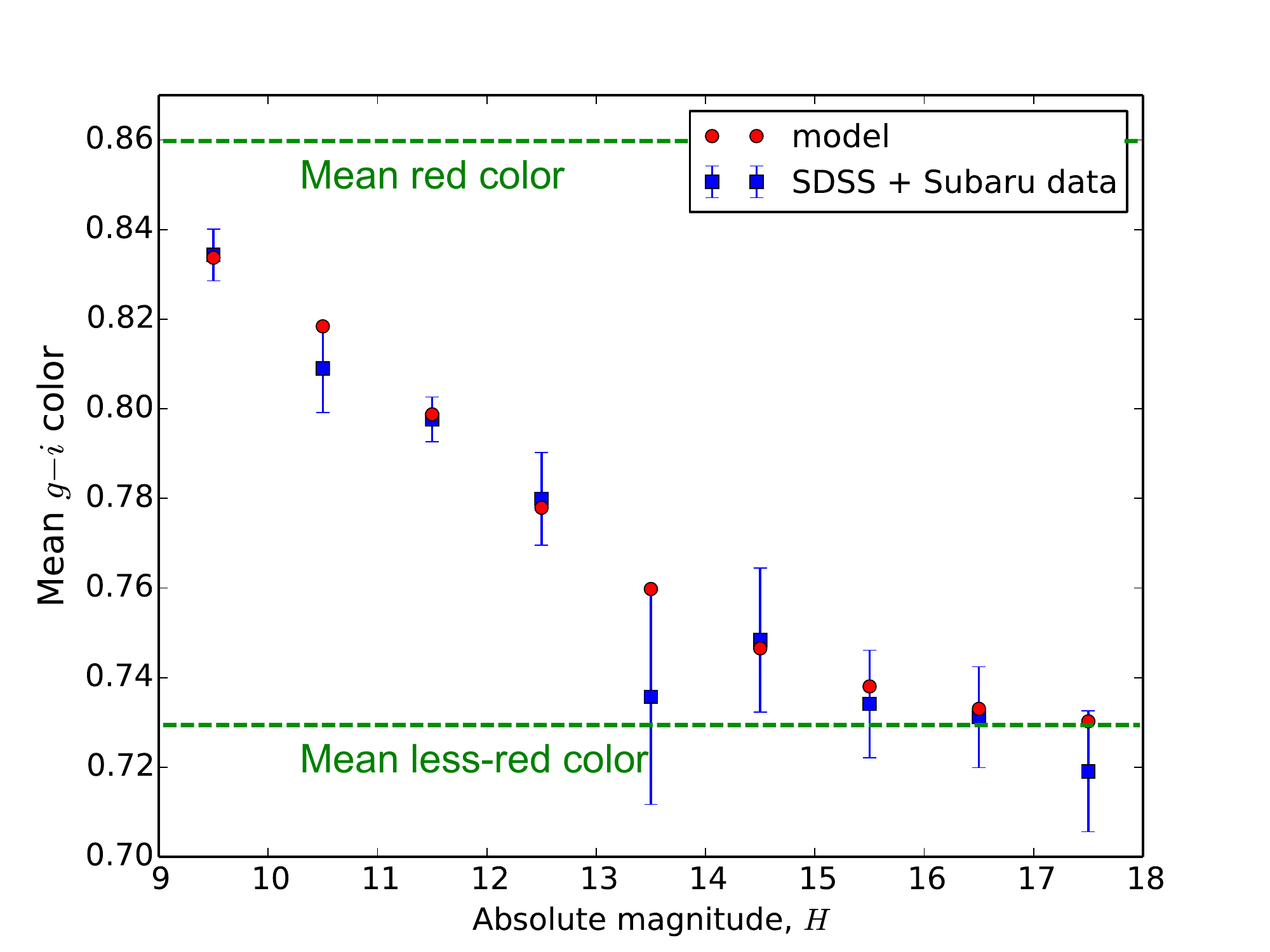}
\end{center}
\caption{Mean $g-i$ colors and corresponding uncertainties for the combined set of Trojans detected in our Subaru survey and L4 Trojans listed in SDSS-MOC4 (blue squares with error bars). Red dots denote the predicted mean $g-i$ color values computed from extrapolation of the best-fit less-red and red population magnitude distributions, assuming mean less-red and red colors of 0.73 and 0.86, respectively. The mean less-red and red $g-i$ colors are indicated by dashed green lines. The monotonic decrease in mean $g-i$ color indicates an increasing fraction of less-red Trojans with decreasing size. The agreement between the extrapolated values and the measured ones suggests that the best-fit color-magnitude distributions derived from bright catalogued Trojans likely extend throughout the magnitude range studied by our Subaru survey.}\label{colortrend}
\end{figure}

Instead, we considered bulk properties of the distribution. In particular, we calculated the mean $g-i$ color as a function of $H$ magnitude for the combined set of SDSS-MOC4 Trojans and faint Trojans from our Subaru observations in order to assess whether the resulting trend is consistent with extrapolation of the best-fit color magnitude distributions obtained previously for the brighter objects (i.e., curves in Figure~\ref{colordist}). Here, we assumed that the mean colors of the two color populations are invariant across all magnitudes. The mean $g-i$ color and uncertainty in the mean for the data were computed in 1~mag bins and are plotted in Figure~\ref{colortrend} in blue. We see that the mean color is consistent with a monotonically-decreasing trend with increasing magnitude, or equivalently, decreasing size. To derive the extrapolated mean color values from the best-fit color magnitude distributions, we calculated the  expected mean color at each bin magnitude as a weighted mean $c'(H)=(\mu_{1}\times\Sigma^{LR}(H)+\mu_{2}\times\Sigma^{R}(H))/(\Sigma^{LR}(H)+\Sigma^{R}(H))$, where $\mu_{1,2}$ are the mean $g-i$ colors of the less-red and red populations, respectively, as derived previously from fitting the color distribution of SDSS-MOC4 Trojans. The functions $\Sigma^{LR}(H)$ and $\Sigma^{R}(H)$ are the best-fit differential magnitude distributions for the less-red and red populations presented earlier. The resulting model mean color values are denoted in Figure~\ref{colortrend} by red dots.

The extrapolated mean colors show very good agreement with the mean colors derived from the data. This suggests that the best-fit magnitude distributions for the less-red and red color populations shown in Figure~\ref{colordist} continue past the limiting magnitude of the MPC data analysis ($H=12.3$) and likely extend throughout most of the magnitude range covered by our Subaru observations. We conclude that the fraction of less-red objects in the overall L4 Trojan population increases steadily with increasing magnitude (decreasing size), and that L4 Trojans fainter than $H\sim 16$, or equivalently, smaller than $\sim$4~km in diameter, are almost entirely comprised of less-red objects.

\section{Discussion}\label{sec:discussion}
The analysis of faint Trojans detected in our Subaru survey offers the most complete picture of the L4 Trojan population to date, refining the known absolute magnitude distribution over the entire range from $H=7.2$ to $H=16.4$ and providing the distribution of Trojan colors down to kilometer-sized objects. The most notable features in the total magnitude distribution (Figure~\ref{total}) are the two slope transitions at $H\approx 8.5$ and $H\approx 15.0$. Such breaks in the power-law shape are common features in the magnitude distributions of small body populations throughout the Solar System and are generally attributed to collisional evolution \citep[see, for example, the review by][]{durda}. Previous studies of the Trojans' collisional history sought to explain the observed bright-end break at $H\approx 8.5$ using collisional modeling \citep[e.g.][]{wong,marzari,deelia} and found that, given the very low intrinsic collisional probability of the Trojan clouds, the bright-end slope transition is best reproduced by assuming that a break was present at the time when the Trojans were emplaced in their current location. In other words, the collisional activity among the large Trojans over the past $\sim$4~Gyr is likely not sufficient to have produced the bright-end break at $H\approx 8.5$ starting from a single power-law slope. We propose that it is the faint-end break at $H\approx 15.0$ that represents the transition to the part of the Trojan population that has reached collisional equilibrium since emplacement; objects brighter than the faint-end break have not reached collisional equilibrium and therefore largely reflect the primordial magnitude distribution of the Trojans at the time of capture by Jupiter.

Turning to the color distributions, we recall that the magnitude distribution fits we obtained for the less-red (LR) and red (R) color populations through $H=12.3$ (see Figure~\ref{colordist}) show highly distinct slopes for objects fainter than the bright-end break ($0.33^{+0.04}_{-0.03}$ for the R population, and $0.61^{+0.07}_{-0.06}$ for the LR population). This indicates that the fraction of LR objects in the overall population increases with increasing magnitude. In Section~\ref{subsec:colordist}, we found that this general trend continues throughout the magnitude range covered in our Subaru survey. As was done in \citet{wong} based on the color-magnitude analysis of L4 and L5 Trojans listed in the Minor Planet Center catalog, we posit that the LR vs. R magnitude distribution slope discrepancy can be explained if R objects convert to LR objects upon collision. Such a process would naturally account for the relative flattening the R population's magnitude distribution slope and the simultaneous steepening of LR population's magnitude distribution slope. 

The R-to-LR conversion model assessed here suggests that the LR and R Trojans have similar interior compositions, with the difference in color confined to the exposed surface layer. Current models of Solar System formation and evolution indicate that the Trojans may have been sourced from the same body of material as the KBOs, located in the region of the disk beyond the primordial orbit of Neptune \citep{morbidelli}. Recent observational studies of KBOs have revealed that the Kuiper belt is comprised of several sub-populations, among which are the so-called ``red'' and ``very red'' small KBOs \citep{fraser,peixinho}. \citet{brown} hypothesized that this color bimodality may be attributable to the wide range of heliocentric distances at which the KBOs formed. In this scenario, all of these objects were accreted from a mix of rock and volatile ices of roughly cometary composition. Immediately following the dissipation of the primordial disk, the surface ices on these bodies began sublimating from solar irradiation, with the retention of a particular volatile ice species on an object's surface being determined primarily by the temperature of the region where the object resided: Objects located at greater heliocentric distances would have retained that ice species on their surfaces, while those that formed at lesser heliocentric distances would have surfaces that were completely depleted in that ice species. \citet{brown} proposed that the continued irradiation of volatile ices led to a significant darkening of the surface and the formation of a robust irradiation mantle, which served to protect ices in the interior from sublimating away. The precise effect of irradiation on the surface is likely dependent on the types of volatile ices retained on the surface. Therefore, the presence or absence of one particular volatile ice species may be the key factor in producing the observed color bimodality in the small KBOs. Specifically, objects that retained that volatile ice species on their surfaces formed a ``very red'' irradiation mantle, while those that lost that volatile ice species from their surfaces formed a ``red'' irradiation mantle.

\citet{wong} suggested that the LR and R Jupiter Trojans may have been drawn from the same two sources as the ``red'' and ``very red" KBOs, respectively. Upon a catastrophic impact, the irradiation mantle on a Trojan's surface would disintegrate and any exposed volatile ices in the interior would sublimate away within a relatively short timescale, leaving behind collisional fragments comprised primarily of water ice and rock. Without the differing collection of volatile ices on the surface to distinguish them, the fragments of LR objects and R objects would be spectroscopically identical to each other. Subsequent irradiation of these pristine fragments would raise the spectral slope slightly, but not to the same extent as would result if volatile ices were retained on the surface. All collisional fragments, regardless of the surface color of their progenitor bodies, would eventually attain the same surface color, which would be relatively less-red when compared to the color of R Trojans. As a consequence, collisional evolution of the Trojan population since emplacement would have gradually depleted the number of R Trojans while simultaneously enriching the number of LR Trojans.

To assess the hypotheses mentioned above, we ran a series of numerical simulations to model the collisional evolution of the L4 Trojan population since emplacement, following the methodology used in \citet{wong}. Given the low intrinsic collisional probability of Trojans, we defined the initial magnitude distribution as a broken power law of the form described in Eq.~\eqref{distribution} with a bright-end distribution identical to that of the currently-observed L4 population ($\alpha_{1} = 0.91$, $H_{0} = 7.22$, and $H_{b} = 8.46$). We varied the initial faint-end slope $\alpha_{2}^{*}$ across different trials. The initial population for each trial consisted of objects with absolute magnitudes in the range $H=7\rightarrow 30$, divided into 75 logarithmic diameter bins. 

We constructed the initial color populations by taking constant fractions of the total initial population across all bins. The initial R-to-LR number ratio, $k$, ranged from 4 to 6, in increments of 0.5. The collisional evolution was carried out over 4 Gyr in 100000 time steps of length $\Delta t= 40000$. At each time step, the expected number of collisions $N_{\mathrm{coll}}$ between bodies belonging to any pair of bins is given by
\begin{equation}N_{\mathrm{coll}}=\frac{1}{4}\langle P\rangle N_{\mathrm{tar}}N_{\mathrm{imp}}\Delta t (D_{\mathrm{tar}}+D_{\mathrm{imp}})^{2},\end{equation}
where $N_{\mathrm{tar}}$ and $N_{\mathrm{imp}}$ are the number of objects in a target bin with diameter $D_{\mathrm{tar}}$ and an impactor bin with diameter $D_{\mathrm{imp}}$, respectively; $\langle P\rangle = 7.79 \times 10^{-18}~\mathrm{yr}^{-1}~\mathrm{km}^{-2}$ is the intrinsic collision probability for Trojan-Trojan collisions  calculated by \cite{delloro} for L$_{4}$ Trojans. For objects with diameter $D_{\mathrm{tar}}$, there exists a minimum impactor diameter  $D_{\mathrm{min}}$ necessary for a shattering collision.  $D_{\mathrm{min}}$ is defined as \citep{bottke}
\begin{equation}D_{\mathrm{min}}=\left(\frac{2Q_{D}^{*}}{V_{\mathrm{imp}}^{2}}\right)^{1/3}D_{\mathrm{tar}},\end{equation}
where $V_{\mathrm{imp}}=4.66~\mathrm{km}~\mathrm{s}^{-1}$ is the L$_{4}$ impact velocity calculated by \cite{delloro}, and $Q_{D}^{*}$ is the collisional strength of target. In our algorithm, we utilized a size-dependent strength scaling law based off one used by \cite{durda} in their treatment of collisions among small main-belt asteroids:
\begin{equation}\label{scaling}Q_{D}^{*} = c\cdot 10\cdot(155.9D^{-0.24} + 150.0D^{0.5} + 0.5D^{2.0})~\mathrm{J\,kg}^{-1}.\end{equation}
Here, we included a normalization parameter $c$ to adjust the overall scaling of the strength; $c$ varied in increments of 0.5 from 1 to 10 in our test trials. The strength scaling model used here has a transition from a gravity-dominated regime for large objects to a strength-dominated regime for smaller objects. For large Trojans, the collisional strength increases rapidly with increasing size, since the impact energy required to completely shatter a large object is primarily determined by the escape velocity of collisional fragments. At smaller sizes (below about 1 km in diameter), the intrinsic material strength of the target becomes the dominant factor; here, smaller objects tend to have fewer cracks and defects than larger objects, and therefore the collisional strength increases with decreasing size.

\begin{figure}[b!]
\begin{center}
\includegraphics[width=9cm]{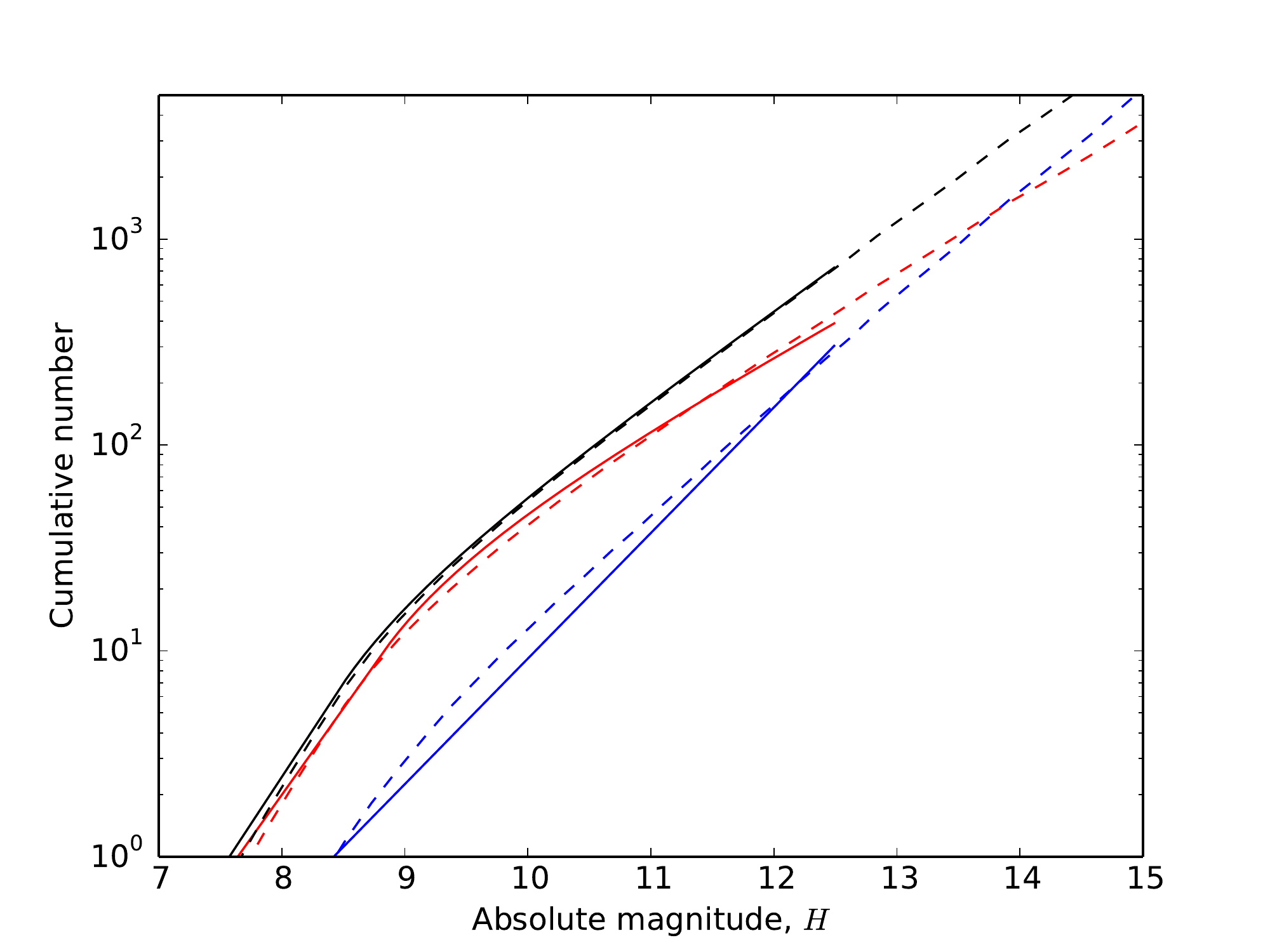}
\end{center}
\caption{Comparison between the results from the best test run of our collisional simulation (dotted lines) and the observed L4 Trojan magnitude distributions (solid lines). The initial total magnitude distribution for the best test run is a broken power law with $\alpha_{1}=0.91$, $\alpha_{2}^{*} = 0.44$, $H_{0}=7.22$, $H_{b} = 8.46$, an initial R-to-LR number ratio $k = 5.5$ and a strength normalization factor $c=1.0$. Black, red, and blue colors indicate the magnitude distribution of the total, red, and less-red populations, respectively. The consistency of the model results and the best-fit distributions demonstrates that the proposed conversion of R objects to LR fragments upon collision is capable of explaining the different shapes of the LR and R magnitude distributions.}\label{simulations}
\end{figure}

Our model computed the collisional evolution of the two color populations separately. At each time step, the simulation considered the number of collisions between objects of the same color, as well as collisions involving objects of different colors. The conversion of red objects to less-red fragments through shattering was modeled by placing all collisional fragments into less-red bins, regardless of the color of the target or the impactor. After running simulations for various values of the parameters ($\alpha_{2}^{*}$, $k$, $c$), we found that a large number of test runs yielded final total and color magnitude distributions that were consistent with the best-fit distributions of catalogued Trojans presented in Section 3. To determine which run best reproduced the observed distributions, we compared the simulation results directly with the best-fit distribution curves and minimized the chi-squared statistic. Here, $\chi^{2}$ was computed as the sum of $\chi^{2}$ values for the total, LR, and R magnitude distributions. The test run that resulted in the best agreement with the data has an initial total distribution with faint-end slope $\alpha_{2}^{*}=0.44$, a strength scaling parameter $c = 1.0$, and an initial R-to-LR number ratio $k = 5.5$. Plots comparing the final simulated distributions from this test run to the best-fit distribution curves are shown in Figure~\ref{simulations}. It is important to note that the simulated total magnitude distribution from the collisional model is not sensitive to the R-to-LR conversion, and the ability of our simulations to reproduce the observed total magnitude distribution holds regardless of any assumptions made about the nature of the color populations.

The similarity between the best initial test distribution and the current total magnitude distribution  reaffirms the conclusion of previous studies that collisions have not played a major role in shaping the magnitude distribution of large Trojans since emplacement. Meanwhile, the R-to-LR collisional conversion model yields simulated final color magnitude distributions that match the best-fit color magnitude distributions of catalogued Trojans well. Furthermore, we computed the expected trend in mean $g-i$ color from the simulated color magnitude distributions through the magnitude region spanning the Subaru Trojan data and found that the trend is consistent with the measured mean $g-i$ colors from the data (as shown in Figure~\ref{colortrend}).

\begin{figure}[t!]
\begin{center}
\includegraphics[width=9cm]{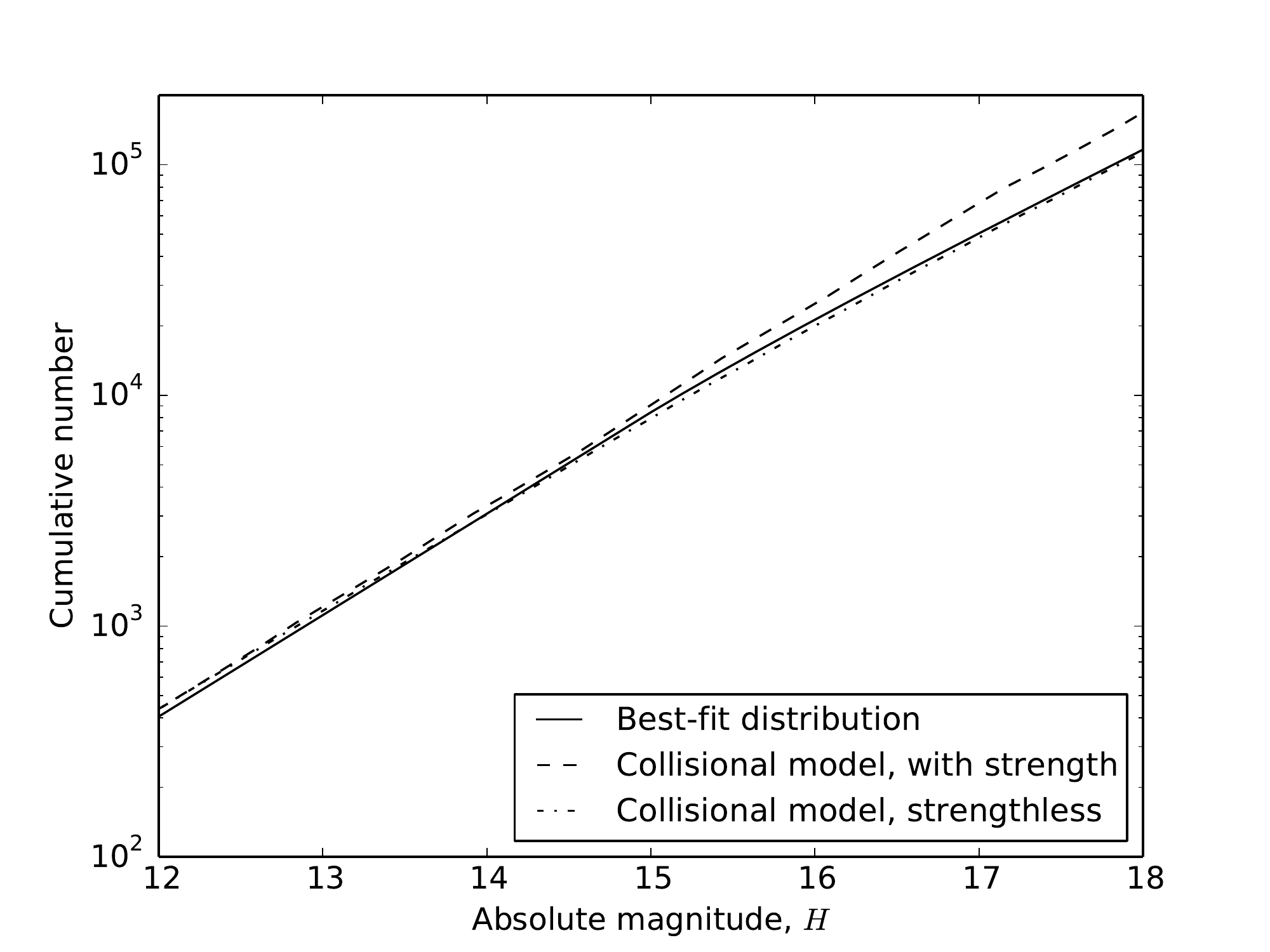}
\end{center}
\caption{Comparison of the best-fit cumulative magnitude distribution curve from the survey data (solid line) with the final predicted distribution from two collisional simulation runs: [1] test run assuming bodies with collisional strength given by Eq.~\eqref{scaling} (dashed line) and [2] test run assuming strengthless bodies with collisional strength scaling given by Eq.~\eqref{scaling2} (dot-dash line). The better agreement of the latter suggests that Trojans have very low material strength, similar to comets.}\label{collcompare}
\end{figure}

To determine whether the collisional model can reproduce the faint-end break at $H\approx 15.0$, we compared the simulated final magnitude distribution from the best test run with the observed magnitude distribution of faint Trojans from our Subaru data (Figure~\ref{collcompare}). All of the simulation test runs predict a break at around $H=14-15$; however, the slope that the simulated distributions roll over to is almost identical to the slope ahead of the break. In the case of the best test run, the faint-end rollover in the simulated final magnitude distribution is barely discernible. In terms of collisional equilibrium, this means that the predicted equilibrium slope is around $\alpha_{\mathrm{eq}}\sim 0.43$. Meanwhile, the actual faint-end slope derived from fitting the Subaru data is somewhat shallower ($\alpha_{2}=0.36^{+0.05}_{-0.09}$).

One possible explanation for this discrepancy is that the collisional strength of Trojans may not be well-described by the strength model defined in Eq.~\eqref{scaling}. To explore this possibility, we considered the case where Trojans have negligible material strength and are loosely-held conglomerates of rock and ice similar to comets. We modified the collisional strength scaling relation to incorporate only the effect of self-gravity by removing the transition to a strength-dominated regime that we included in the previous model. The new strength formula is given by:
\begin{equation}\label{scaling2}Q_{D}^{*} = c\cdot 5D^{2.0}~\mathrm{J\,kg}^{-1}.\end{equation}
Rerunning the collisional simulations with this new collisional strength scaling, we established a new best test run for strengthless bodies with $\alpha_{2}^{*}=0.45$, $k = 5.5$, and $c = 0.5$. The simulated color magnitude distributions in the strengthless case were found to be generally consistent with the ones produced in the original non-strengthless model and likewise predict the observed trend in mean $g-i$ color through $H\sim 18$. The simulated final magnitude distribution from this model in the vicinity of the faint-end break is shown in Figure~\ref{collcompare}. We can see that the predicted distribution for strengthless Trojans provides a better match to the data than the previous case of non-strengthless bodies. This indicates that Trojans may have very low material strength, which would make them more comparable to comets than to main belt asteroids and lend support to the hypothesis presented earlier that the Jupiter Trojans formed in the primordial trans-Neptunian region and were later scattered inward during a period of dynamical instability.

\section{Conclusion}
We detected 557 Trojans in a wide-field survey of the leading L4 cloud using the Suprime-Cam instrument on the Subaru Telescope. All objects were imaged in two filters, and the $g-i$ color was computed for each object. After removing objects imaged during bad seeing and establishing a limiting magnitude of $H=16.4$, we computed the best-fit curves describing the overall magnitude distribution. In addition, we examined the distribution of $g-i$ colors for the faint objects detected by our survey and compared it to an extrapolated model based on the magnitude distributions of bright, catalogued objects in the less-red and red Trojan populations. The color-magnitude distribution analysis was supplemented by collisional simulations, from which we made predictions about the formation and evolution of the Trojan population. The main results are summarized below:
\begin{itemize}
\item The overall magnitude distribution of L4 Trojans is described by three power-law slopes: The distribution of the brightest objects follows a power law slope of $\alpha_{0}=0.91^{+0.19}_{-0.16}$. At intermediate sizes, the magnitude distribution rolls over to a slope of $\alpha_{1}\sim 0.44$. Finally, the faintest objects are characterized by an even shallower magnitude distribution slope of $\alpha_{2}=0.36^{+0.05}_{-0.09}$. These three regions are separated by rollovers in the magnitude distribution located at $H'_{b}=8.46^{+0.49}_{-0.54}$ and $H_{b}=14.93^{+0.73}_{-0.88}$, which correspond to objects with diameters of $135^{+38}_{-27}$~km and $7^{+3}_{-2}$~km, respectively.
\item The faint-end break in the overall Trojan magnitude distribution at $H\sim 15$ is reproduced by our collisional simulations and indicates the transition between objects that have not experienced significant collisional evolution and objects that have achieved collisional equilibrium.
\item The shallow faint-end slope ($\alpha_{2}=0.36^{+0.05}_{-0.09}$) is consistent with Trojans having very low material strength, similar to comets.
\item The mean $g-i$ color of Trojans follows a general decreasing trend with increasing magnitude, or equivalently, decreasing size. At faint magnitudes, this trend is consistent with the extrapolation of magnitude distribution fits computed for bright objects in the less-red and red populations. Less-red objects dominate among objects smaller than $\sim$5~km in diameter.
\item The discrepant best-fit slopes of the color-magnitude distributions for objects smaller than $\sim$50~km and the monotonically-decreasing trend in mean $g-i$ color with decreasing size are consistent with the conversion of red objects to less-red fragments upon collision.
\end{itemize}

\section*{Acknowledgments}
This research was supported by Grant NNX09AB49G from
the NASA Planetary Astronomy Program and by the Keck Institute for Space Studies. The authors also thank an anonymous reviewer for constructive comments that helped to improve the manuscript.

\end{document}